\documentclass[prc,preprint,aps,showpacs,nofootinbib,12pt]{revtex4-1}
\newcommand{\bea}{\begin{eqnarray}}
\newcommand{\eea}{\end{eqnarray}}
\usepackage{amssymb}
\usepackage{graphics}
\begin{document}

\title{'Running' under tight constraints in pionless effective field theory}
\author{Jun-Jun L\"u\footnote{Present address: Jiujiang TongWen Middle School, Jiujiang, Jiangxi Province 332000, China.}, Ji-Feng Yang\footnote{Corresponding author.}}
\affiliation{School of Physics and Electronic Science, East China Normal University, Shanghai 200241, P.R. China}
\begin{abstract}
The contents of renormalization group invariance and equations under tight constraints are explored and demonstrated with closed-form on-shell $T$ matrices of pionless effective field theory for nuclear forces right within the effective field theory philosophy. The 'running' couplings under such tight constraints are presented in $^1S_0$ and uncoupled $P$ channels up to truncation order $\mathcal{O}(Q^4)$. Some linear relations are exposed and in turn employed in the pursuit of nonperturbative 'running' solutions which serves as an alternative choice without resorting to special prescription and additional operations or treatments. The utility of such 'running' behaviors inherent in the closed-form $T$-matrices of pionless effective field theory is remarked and a number of important issues related to effective field theory constructed with various truncations are interpreted or discussed from the underlying theory perspective.
\end{abstract}
\pacs{3.65.Nk;21.45.Bc;11.10.Gh}
\maketitle

\section{introduction}In literature, there have been various forms of renormalization group equation that describe running behaviors in the framework of field theories. In a sense, the known renormalization group equations could be cast into the following two categories: (a) renormalization group equations derived from the standard algorithm of perturbative renormalization\cite{Stuek,Gell,Callan,Syman,tHooft,Weinb}; (b) Wilsonian renormalization group equation\cite{Wilson} and exact renormalization group equation\cite{Weg,Nicoll,Polch,Wett,Ell,BBreview} that work essentially in nonperturbative regime. In this report, we wish to explore another avenue of renormalization group equation: nonperturbative running of in the presence of tight constraints, which is imposed by the closed-form of $T$ matrices obtained from Lippmann-Schwinger equations (LSE) or Schr\"odinger equations in a general parametrization of momentum integrals. Specifically, the objects under consideration are the closed-form $T$ matrices for $NN$ scattering. Earlier studies of conventional renormalization group equation within LSE could be found in Refs.\cite{Frederico1,Frederico2}.

Our studies originate from the effective field theory approach for nuclear force, for reviews, see\cite{review1b,review2a,reviewEM,review2b,review5}. For energies much below pion mass, one could simply work with the simpler pionless effective field theory or contact potentials so that closed-form solutions to Lippmann-Schwinger equations are feasible\cite{YH,JPA,AOP}. However, these closed-form $T$ matrices imposes tight constraints that precludes the conventional renormalization program (established in perturbation theory) from working, hence are usually discarded or circumvented.

Nevertheless, an elaborate analysis of the closed-form $T$-matrices showed that the tight constraints could actually be exploited to arrive at a novel nonperturbative scenario for effective field theory renormalization\cite{AOP}. In such scenarios, not all parameters from effective field theory loops could be absorbed into the effective field theory couplings, some of them must be treated as physical and hence renormalization group invariant parameters to be fixed independently\cite{AOP}. The conceptual foundation of such treatment just lies in the underlying theory perspective: effective field theory is only a simple description of certain phenomena, it is conceivable that some ingredients of the regular properties of underlying theory are not adequately described by effective field theory couplings but show up through effective field theory (loop) integrals in nonperturbative contexts with various truncations and usually contaminated by UV divergences. Hence the real issue in effective field theory is to 'fix' the effective field theory loop integrations in ways compatible with underlying theory principles as far as possible. Such underlying theory perspective naturally allows us to employ a general parametrization of the effective field theory integrals (and the associated subtraction), which are immune from resorting to special prescription like power divergence subtraction (PDS) and additional treatments\cite{KSW1,KSW2,Gege}.

In a sense, the tight constraints imposed by the closed-form $T$ matrices may render part of the parameters from effective field theory loop integrals encrypted with some physical contents and the running behaviors of effective field theory couplings quite different from that in perturbative regimes. Thus, examining the nonperturbative running couplings and their properties may yield us further conceptual gains. In spite that our discussion is given in the much simpler pionless effective field theory, the concepts and scenarios discussed here could in principle be extended to other effective field theories. As we employ a general parametrization of effective field theory loop integrals, which is a crucial component of our approach, our discussion should be useful for all the physical studies that are beset with ultra-violet divergences in nonperturbative contexts, where conventional programs are actually infeasible. In other words, we employ a simple and yet nontrivial setting to demonstrate alternative scenario of renormalization in nonperturbative contexts and its utilities, which has been overlooked so far in most literature.

This report is organized as below: In Sec. II, we present the results of the closed-form on-shell $T$ matrices in $^1S_0$ channel and the uncoupled $P$ channels at truncation order $\mathcal{O}(Q^4)$ and some linear relations among the factors for parameterizing the closed-form $T$ matrices; Sec. III is devoted to the plausible reasonings for the issue of effective field theory renormalization and renormalization group invariance in nonperturbative contexts, which in turn leads us to the version of renormalization group equation in the presence of tight constraints imposed by closed-form $T$ matrices; In Sec. IV, we utilize the tight constraints to obtain nonperturbative running couplings in $^1S_0$ channel and the uncoupled $P$ channels, and explore for their properties and implications; The summary is given in Sec. V.
\section{Closed-form $T$ matrices in pionless effective field theory for $NN$ scattering}
\subsection{Parametrization of the on-shell $T$}In pionless effective field theory, the $NN$ interaction becomes contact one. The contact potential truncated at order $\mathcal{O}(Q^4)$ in $^1S_0$ channel reads:\bea V_{^1S_0}=C_0+C_2\left(q^2+{q^{\prime}}^2\right)+C_4\left(q^4+{q^{\prime}}^4\right)+\tilde{C}_4q^2{q^{\prime}}^2,\eea from which it is easy to read off the potential truncated at orders $\mathcal{O}(Q^0)$ and $\mathcal{O}(Q^2)$. The closed-form $T$ matrices for such contact potentials could be found by solving the LSE via the trick employed in Ref.\cite{Maryland} and elaborated in Ref.\cite{YH} where the convolution integral is parameterized in a general manner. Then, the on-shell $T$ matrix for $^1S_0$ channel at order $\mathcal{O}(Q^4)$ reads\cite{YH},\bea\label{Tn3}&&\frac{1}{T(p)}=\mathcal{I}_0+\frac{N_{S;0}+N_{S;1}p^2+N_{S;2}p^4}{D_{S;0}+D_{S;1}p^2+D_{S;2}p^4+D_{S;3}p^6},\\&&\mathcal{I}_0 \equiv J_0+i\frac{M_N}{4\pi}p,\ p\equiv\sqrt{M_NE}.\eea The detailed expressions of $[N_{\cdots}]$ and $[D_{\cdots}]$ are listed in \ref{appA}, where we also give the definition of the prescription-dependent parameters $[J_{2n+1},n>0]$ that arise from the convolution integration. At this stage they are referred to as a general parametrization of regularization/renormalization.

Putting higher order couplings to zero would yield the corresponding lower order results for $[N_{\cdots}]$ and $[D_{\cdots}]$, for example, putting $C_4=\tilde{C}_4=0$ will lead us back to $\mathcal{O}(Q^2)$:\bea&&\frac{1} {T(p)}=\mathcal{I}_0+\frac{N_{S;0}}{D_{S;0}+D_{S;1}p^2},\\&&N_{S;0}=(1-C_2J_3)^2,\ D_{S;0}=C_0+C_2^2J_5,\ D_{S;1}=2C_2-C_2^2J_3.\eea Putting all except $C_0$ to zero will lead us to $\mathcal{O}(Q^0)$: $1/T(p)=\mathcal{I}_0+1/C_0.$

For the uncoupled $p$-wave channels, we have at order $\mathcal{O}(Q^4)$:\bea&&V_{^{2s+1}P_j}=C_{P;2}qq^{\prime}+C_{P;4}\left(q^2+{q^{\prime}}^2\right)qq^{\prime}\nonumber\\& &\rightarrow\frac{1}{T(p)}=\mathcal{I}_0+\frac{N_{P;0}+N_{P;1}p^2}{D_{P;0}p^2+D_{P;1}p^4}.\eea The detailed expressions for $[N_{\cdots}]$ and $[D_{\cdots}]$ are listed in \ref{appB}. At order $\mathcal{O}(Q^0)$, the $T$ matrix does not exist as $V_{^{2s+1}P_j}=0$. At order $\mathcal{O}(Q^2)$, we have\bea&&\frac{1}{T(p)}=\mathcal{I}_0+\frac{N_{P;0}}{D_{P;0}p^2}, \nonumber\\&&N_{P;0}=1-C_{P;2}J_3,\ D_{P;0}=C_{P;2}.\eea

The off-shell version of $T$ matrix at generic order is more involved and the $\mathcal{O}(Q^4)$ case for $^1S_0$ and uncoupled $P$ channels will be presented in \ref{appC}. Here, to get a rough idea, we list the $\mathcal{O}(Q^2)$ case:\bea^1S_0:&&{T_S(q,q^\prime;p)}=\frac{\tilde{D}_S(q,q^\prime;p)+\check{\delta}_S(q,q^\prime;p)\check{O}_{\texttt{\tiny off}}} {N_{S;0}+\mathcal{I}_0\left(\sum_{i=0}^1D_{S;i}p^{2i}\right)},\quad\check{O}_{\texttt{\tiny off}}\equiv\left(p^2-q^2\right)\left(p^2-{q^\prime}^2\right),\nonumber\\&&\tilde{D}_S (q,q^\prime;p)=V_S(q,q^\prime)+C^2_2\left[J_5+J_3\left(p^2-q^2-{q^{\prime}}^2\right)\right],\quad\check{\delta}_S(q,q^\prime)=-\mathcal{I}_0C^2_2.\\P:&&{T_P(q,q^\prime;p)}=\frac{\tilde {D}_P(q,q^\prime;p)+\check{\delta}_P(q,q^\prime;p)\check{O}_{\texttt{\tiny off}}}{N_{P;0}+\mathcal{I}_0D_{P;0}p^2},\nonumber\\&&\tilde{D}_P(q,q^\prime;p)=V_P(q,q^\prime),\quad \check{\delta}_P(q,q^\prime)=0.\eea It is obvious that the item containing $\check{O}_{\texttt{\tiny off}}$ is a pure off-shell part, and the on-shell momentum or energy enter the game 'everywhere' in the functional form of $T$ matrix, while in the momentum transfer only appear in the numerator. It is a simple exercise to verify that $\tilde{D}_S(p,p;p)= \sum_{i=0}^1D_{S;i}p^{2i}$ at this order. We would also like to note that the off-shell $T$ matrix coincides exactly with the on-shell $T$ matrix only at the leading order in $^1S_0$ channel, making it a very special case.
\subsection{Linear constraints for $[N_{\cdots},D_{\cdots}]$}Interestingly, there exist some linear relations among the factors $[N_{S,\cdots},D_{S,\cdots}]$ at each order of truncation:\bea\label{relation1s0-0}\mathcal{O} (Q^0):&&N_{S;0}=1,\ D_{S;0}=C_0.\\\label{relation1s0-2}\mathcal{O}(Q^2):&&N_{S;0}+D_{S;1}J_3=1,\nonumber\\&&D_{S;0}J_3+D_{S;1}J_5=C_0J_3 +2C_2J_5.\\\label{relation1s0-4}\mathcal{O}(Q^4):&&N_{S;2}+D_{S;3}J_3=0,\ N_{S;1} +D_{S;2}J_3+D_{S;3}J_5=0,\nonumber\\&&N_{S;0}+D_{S;1}J_3+D_{S;2}J_5+D_{S;3}J_7=1,\nonumber\\&&\sum _{n=0}^{3}D_{S;n}J_{2n+3}=C_0J_3+2C_2J_5+(2C_4+\tilde{C}_4)J_7.\eea All these relations could be readily verified. Given these evidences, we are tempted to conjecture that the following linear relations for $^1S_0$ channel at a generic order $\Delta$:\bea\label{NDrelation1s0}\mathcal{O}(Q^\Delta):&&N_{S;k}+\sum_{n=k+1}^{\Delta-1}D_{S;n}J_{2(n-k)+1}=0,\ 1\leq k\leq\Delta-2,\nonumber\\&&N_{S;0}+\sum_{n=1}^{\Delta-1}D_{S;n}J_{2n+1}=1,\nonumber\\&&\sum_{n=0}^{\Delta-1}D_{S;n}J_{2n+3}=\sum_{n=0}^{\Delta/2}\left(a_nC_{2n}+b_n\tilde{C}_{2n} +\cdots\right)J_{2n+3}.\eea Here, $a_n=b_n=\cdots=1$ for diagonal entries and $a_{n}=b_n=\cdots=2$ for off-diagonal entries.

Similarly, for $P$-channels, we have verified up to $\mathcal{O}(Q^4)$ and again conjecture that\bea\label{relationP2}\mathcal{O}(Q^2):&&N_{P;0}+D_{P;0}J_3=1,\nonumber\\&&D_{P;0}J_5 =C_{P;2}J_5.\\\label{relationP4}\mathcal{O}(Q^4):&&N_{P;1} +D_{P;1}J_3=0,\nonumber\\&&N_{P;0}+D_{P;0}J_3+D_{P;1}J_5=1,\nonumber\\&&D_{P;0}J_5+D_{P;1}J_7=C_{P;2}J_5+2C_{P;4}J_7.\\ \label{NDrelationP}\mathcal{O}(Q^\Delta):&&N_{P;k}+\sum_{n=k}^{\Delta-3}D_{P;n}J_{2(n-k)+3}=0,\quad1\leq k\leq\Delta-3,\nonumber\\&&N_{P;0}+\sum_{n=0}^{\Delta-3}D_{P;n}J_{2n+3}=1, \nonumber\\&&\sum_{n=0}^{\Delta-3}D_{P;n}J_{2n+5}=\sum_{n=1}^{\Delta/2}\left(a_nC_{P;2n}+b_n\tilde{C}_{P;2n}+\cdots\right)J_{2n+5}.\eea Again, $a_n=b_n=\cdots=1$ for diagonal entries and $a_{n}=b_n=\cdots=2$ for off-diagonal entries.

We have verified that these relations or identities are also valid at order $\mathcal{O}(Q^6)$ in $^1S_0$ channel. In Sec. IV, we will see that these relations are crucial for finding the solutions of nonperturbative running couplings. At present we could provide neither a rigorous proof nor a sound interpretation of these relations. Further explorations in the future are worthwhile. There might be some intriguing contents in these relations that could be illuminating.

In pionless effective field theory, the contact couplings are dominated by pion-exchange loop diagrams in the range $l\in(m_\pi,M_N)$, such intimate relations between $[C_{\cdots}]$ and $[J_{\cdots}]$ imply that they should come from different 'sides' of the same regularities of a underlying theory, say, covariant chiral perturbation theory, or quantum chromo-dynamics (QCD). It is also obvious that the 'contents' of these relations depend upon specific prescriptions of $[J_{\cdots}]$. They would reduce to very uninteresting and uninformative ones in dimensional schemes like PDS\cite{KSW1,KSW2}. In a sense, regularization scheme really matters in nonperturbative regime, it may reveal or hide some intricacies of the underlying theory, in contrast to the conventional wisdoms.
\section{Renormalization group invariance in nonperturbative regime}
\subsection{Standard renormalization group equation as a 'decoupling theorem' from underlying components}Let us first digress a little on the general form of renormalization group equation from the underlying theory perspective, where the 'corrections' from the 'underlying components' ($\{\sigma\}$ that render an effective field theory well defined in the ultra-violet region) to the canonical scaling laws in effective field theories could be readily interpreted as 'decoupling theorems': The scalings of the 'underlying components' ($\sum_{\sigma}d_{\sigma}\sigma\partial_{\sigma}$) could at most contribute that of local effective field theory (composite) operators $[O_i]$ in the 'decoupling limits' provided that the effective field theory is local and covariant\cite{PLB625}:\bea&&\sum_{\sigma}d_{\sigma}\sigma\partial_{\sigma}\Gamma^{(n)}([p],[g];\{\sigma\})\nonumber\\&&\Rightarrow \sum_{\bar{c}}d_{\bar{c}}{\bar{c}}\partial_{\bar{c}}\Gamma^{(n)}([p],[g];\{\bar{c}\})=\sum_{O_i}\delta_{O_i}I_{O_i}\Gamma^{(n)}([p],[g];\{\bar{c}\}),\eea where $[p]$ and $[g]$ being external momenta and couplings (including masses) in a complete $n$-point function $\Gamma^{(n)}$, $d_{\cdots}$ being the mass dimension, $\{\bar{c}\}$ the constants from the 'decoupling limits'\footnote{In the conventional algorithm of perturbative renormalization, they arise from the subtraction procedure.}, and $\delta_{O_i}$ the anomalous dimension of $O_i$. This somewhat 'primitive' form of renormalization group equation puts renormalizable and nonrenormalizable theories on the same footing and could be readily transformed into various other forms from which we could readily recover some well-known low-energy theorems derived in QCD in renormalized form as natural corollaries\cite{PLB625,JPA40}. In covariant perturbation theory, the 'anomalous' contributions of $\{\bar{c}\}$ to the scaling laws come from the logarithmic terms like $\delta_{O_i}\ln\frac {\bar{c}}{m}$ ($d_{\bar{c}}=1=d_m$, $m\in[g]$, i.e., $m$ is a mass in effective field theory) that arise from loops.

However, when various truncations are employed, the standard renormalization group equation may no longer be ensured. Furthermore, sticking to the standard renormalization group equation and the associated wisdom would lead us nowhere in front of the tight constraints that arise from the combination of truncations and nonperturbative contexts. For example, for the on-shell closed-form $T$ matrices given in Sec. II.A., the scaling law simply read\bea\label{RGET}\left\{p\partial_p+\sum_{C_{\cdots}}d_{C_{\cdots}}C_{\cdots}\partial_{C_{\cdots}} +\sum_{J_{\cdots}}d_{J_{\cdots}}J_{\cdots}\partial_{J_{\cdots}}-2\right\}{T}(p;[C_{\cdots}];[J_{\cdots}])=0,\eea where $\sum_{C_{\cdots}}d_{C_{\cdots}}C_{\cdots}\partial_{C_{\cdots}}$ denotes the contribution from contact couplings and $\sum_{J_{\cdots}}d_{J_{\cdots}}J_{\cdots}\partial_{J_{\cdots}}$ denotes that from loop integrals (which should correspond to $\sum_{\bar c}d_{\bar{c}}{\bar{c}}\partial_{\bar{c}}$). Since $[C_{\cdots}]$ and $[J_{\cdots}]$ are all power like dimensional constants and highly intertwined on the same footing in the homogeneous polynomial factors ($[N_{\cdots}]$ and $D_{\cdots}]$) in the closed-form $T$ matrices, the scaling of $[J_{\cdots}]$ could no longer be simply cast into scaling anomalies in terms of local operators and absorbed into contact couplings. Actually, there is in general a mismatch between $[C_{\cdots}]$ and $[J_{\cdots}]$ just due to truncation\cite{JPA}.

Such closed-form solutions would be usually deemed a disaster and discarded or circumvented at all. Here, we take this difficulty as a motivation to reexamine the whole issue from general principles so as to make sense out of the tight constraints. In our view, the best choice is to resort to the underlying theory perspective view of effective field theory renormalization in nonperturbative contexts so as to turn the tight constraints into virtues to be exploited, which will be discussed and explicated below in Sec. III.B.
\subsection{Nonperturbative scenario of effective field theory renormalization in underlying theory perspective}Our take on the issue is based on the following observations\cite{YH,JPA,AOP}:

First, in the underlying theory perspective ultra-violet divergences arise in effective field theory as the effective field theory projection operation $\breve{{\mathcal{P}}}_{\texttt{\tiny EFT}}$ does not commutate with loop integrations\cite{YH},\bea\texttt{CT}\equiv[\breve{{\mathcal {P}}}_{\texttt{\tiny EFT}},\int\!d^Dl] \neq0.\eea Then subtraction automatically follows in each loop as a rearrangement of this commutator ($\texttt{CT}$),\bea\breve{{\mathcal{P}}}_{\texttt{\tiny EFT}}\!\!\int\!d^Dl\ \underbrace{[f(l,\cdots)]}_{UT}=\int\!d^Dl\ \underbrace{\breve{{\mathcal{P}}}_{\texttt{\tiny EFT}}\!\left[f(l,\cdots)\right]}_{EFT}+\underbrace{\texttt{CT}\left[f(l,\cdots)\right]} _{counterterm},\eea giving rise to the parameters (here, the $[J_{\cdots}]$) to be 'fixed'.

Second, there are intrinsic mismatches between the parameters and the effective field theory couplings in the closed-form $T$ matrices\cite{JPA}, the 'matched' ones could be absorbed into the effective field theory couplings $[C_{\cdots}]$ and make the latter 'run', while the 'unmatched' ones are separately constrained to be renormalization group invariant or physical, giving rise to the following scenario for effective field theory renormalization: $$\mathcal{S}\equiv[C_{\cdots}(\mu)]\oplus[J_{\cdots}^{\texttt {\tiny(phys)}}, \tilde{J}_{\cdots}(\mu)]=[C_{\cdots}(\mu),\tilde{J}_{\cdots}(\mu)]\oplus[J_{\cdots}^{\texttt{\tiny(phys)}}]$$ with $\mu$ a running scale.

Consequently, $[C_{\cdots}(\mu)]$ and $[\tilde{J}_{\cdots}(\mu)]$ must conspire in the remaining renormalization group invariants and join with $[J_{\cdots}^{\texttt{\tiny(phys)}}]$ to parameterize the closed-form $T$ matrices\cite{AOP}. Therefore, we need to find these renormalization group invariants and then solve the running couplings in terms of these renormalization group invariants and running parameters $[\tilde{J}_{\cdots}(\mu)]$, which are the subjects of the next two subsections. This conception of renormalization in front of tight constraints deviates from the standard wisdom established in perturbative contexts.
\subsection{Renormalization group invariance of the shape of on-shell $T$}First, the dependence of an on-shell $T$ matrix on $p$ (the functional shape) is physical since it is related to the phase shift in the following way in channel $L$:\bea\Re \left\{-\frac{4\pi p^{2L}}{M_NT(p)}\right\}=p^{2L+1}\cot\delta_L(p).\eea In the present work, the functional shape of the closed-form on-shell $T$ matrices are completely encoded in the ratios like $[N_{L;i}/N_{L;0}, D_{L;j}/N_{L;0}]$ and the parameter $J_0$, so these ratios must be physical and hence renormalization group invariant in general sense:\bea&&\left\{\frac{\sum_{i}\left(N_{L;i}/N_{L;0}\right)p^{2i}}{\sum_j\left(D_{L;j}/N_{L;0}\right)p^{2j}}+J_0p^{2L}\right \}_{{RG\ inv}}\nonumber\\&&\label{RG-ND-npt}\Longrightarrow\delta_{{RG}}(J_0)=0,\ \delta_{{RG}}\left(\frac{N_{L;i}}{N_{L;0}}\right)=0,\ \delta_{{RG}}\left(\frac{D_{L;j}}{N_{L;0}} \right)=0,\ \forall i,j.\eea Here, '$\delta_{{RG}}$' denotes the variations in renormalization prescriptions. In case of infinitesimal variations, they are nothing else but the homogeneous renormalization group equations satisfied by these (physical) ratios, a fact long established in renormalization theory.

In a complete formulation in terms of underlying theory, renormalization group invariance should read $\delta_{RG}\left\{\cdots\right\}=\left(\sum_{\sigma}d_{\sigma}\sigma \partial_{\sigma}-\sum_{O_i}\delta_{O_i}I_{O_i}\right)\left\{\cdots\right\}=0,$ while here in pionless effective field theory, it reads $$\delta_{{RG}}=\sum_{\tilde{J}_{\cdots}}d_{\tilde{J}_{\cdots}}\tilde{J}_{\cdots}(\mu)d_{\tilde{J}_{\cdots}(\mu)}=\sum_{\tilde{J}_{\cdots}}d_{\tilde{J}_{\cdots}}\tilde{J}_{\cdots}(\mu) \partial_{\tilde{J}_{\cdots}(\mu)}+\sum_{C_{\cdots},\tilde{J}_{\cdots}}\gamma_{C_{\cdots};\tilde{J}_{\cdots}}C_{\cdots}(\mu)\partial_{C_{\cdots}(\mu)},$$ $$\gamma_{C_{\cdots};\tilde{J}_{\cdots}}\equiv\partial{C}_{\cdots}/\partial{\tilde{J}_{\cdots}},$$ where the contributions from $[J_{\cdots}^{\texttt{\tiny(phys)}}]$ must be excluded due to tight constraints. By contrast, in Wilsonian renormalization group equations, $\delta_{{RG}}$ is simply implemented as $\Lambda{d}_\Lambda$ ($\Lambda$ denoting the cutoff scale), which is okay for perturbative issues. However, in presence of tight constraints it might be problematic. The reason goes as below: Using a universal cutoff $\Lambda$, each of $[J_{\cdots}]$ is $\Lambda$-dependent. However, the tight constraints require some of them, say, $[J_{\cdots}^{\texttt{\tiny(phys)}}]$, to be 'physical' or renormalization group invariant. As the variation with $\Lambda$ also induces the changes in $[J_{\cdots}^{\texttt{\tiny(phys)}}]$ (denoted as $[\Lambda{\partial_\Lambda}]_{\texttt{\tiny(Phys)}}$), this component must be excluded to ensure the real renormalization group invariance, i.e., $\delta_{RG}=\Lambda{d_\Lambda}-[\Lambda{\partial_\Lambda}]_{\texttt{\tiny(Phys)}}$. Evidently, it is a mission impossible within the conventional form of Wilsonian renormalization group equations\footnote{Unless some sophisticated ingredients are introduced to alleviate the pressure, see a recent effort in this direction in Ref.\cite{EGeM}.}. One must try to seek for tractable ways to realize the operation $\delta_{RG}=\Lambda{d}_\Lambda-[\Lambda{\partial_\Lambda}]_{\texttt{\tiny(Phys)}}$. Obviously, the general parametrization of loop integrals plays a pivotal role in the foregoing discussion.

Now we should note that $J_0$ becomes renormalization group invariant or physical at a generic truncation order. It corresponds to the constant part of the fixed-point solution of $NN$ scattering in Ref.\cite{Birse}. As, the ratios that satisfy Eqs.(\ref{RG-ND-npt}) are physical or renormalization group invariant, the effective field theory couplings must 'run' in ways to exactly cancel the running parameters to keep these ratios intact. Then the running couplings could be found reversely from these physical ratios, which is our job to be done in the following sections.
\subsection{Renormalization group invariance of ERE parameters}Since the effective range expansion (ERE) is actually a Taylor expansion around $p=0$, the functional shape of $T$-matrices are also completely encoded in the ERE factors:\bea\label{ERT}\Re\left\{-\frac{4\pi p^{2L}}{M_NT(p)}\right\}=p^{2L+1}\cot\delta_L(p)=-\frac{1}{a}+\frac{1}{2}r_ep^2 +\sum_{k=2}^{\infty}v_k p^{2k},\eea with $a$ and $r_e$ being known as the scattering length and effective range in channel $L$. All the ERE factors are physical observables and hence could serve as the renormalization group invariants for our purpose, as they are rational functions in terms of $[N_{L;i},D_{L;j}]$ and $J_0$.

Actually, one could retrieve the renormalization group invariant ratios defined in Sec. III.C. from the ERE factors through appropriate combinations, at any given order of truncation. To illustrate, we take the $^1S_0$ channel at order $\mathcal{O} (Q^2)$ as a simple but nontrivial example, where\bea&&a^{-1}=\frac{4\pi}{M_N}(N_{S;0}{D_{S;0}}^{-1}+J_0),\ r_e= \frac{8\pi}{M_N}N_{S;0} D_{S;1}D_{S;0}^{-2},\nonumber\\&&v_k=(-1)^{k-1}\frac{4\pi}{M_N}N_{S;0} D_{S;1}^kD_{S;0}^{-k-1},\ k\geq2.\eea It is obvious that the following ratios are also renormalization group invariant:\bea&&\Xi_0\equiv\frac{M_Nr_ev_k}{8\pi v_{k+1}}+\frac{M_N}{4\pi a}=J_0,\ \Xi_1\equiv\frac{8\pi v_{k+1}}{M_Nr_ev_k}=-\frac{D_{S;0}}{N_{S;0}},\nonumber\\ &&\Xi_2\equiv\frac{8\pi v^2_{k+1}}{M_Nr_ev_k^2}=\frac{D_{S;1}}{N_{S;0}},\eea just equivalent to the Eqs.(\ref{RG-ND-npt}) at $\mathcal{O}(Q^2)$. Such kind of solutions are always feasible right due to truncation, which is exploited here rather than circumvented.

We should note in passing that, in the PDS prescription, $J_{\cdots}=0, J_0\neq0$, most of the ERE parameters of the $^1S_0$-channels turn out to be rational functions of the contact couplings like\bea a^{-1}=\frac{4\pi}{M_N}\left(C^{-1}_0+J_0\right),\ r_e=F_2(C_0,C_2),\ v_k=F_k(C_0,[C_{2j}]),k\geq2.\eea One might think that this will lead us back to the Kaplan-Savage-Wise (KSW) running\cite{KSW1,KSW2} in this channel. However, with {\em the combinations} given above, we could always arrive at {\em an alternative solution} that is truly 'nonperturbative' in essence, see Sec. IV.C. This alternative has so far been overlooked in most literature. The well-known KSW running could be recovered for $C_0$ in $^1S_0$ channel, but only at the lowest order of truncation.

We should note in passing that the deductions of the foregoing two subsections (III.C. and III.D.) do not mean the renormalization group invariant ratios are exact physical numbers, but that allowing for systematic effective field theory corrections according to normal power counting rules, which will be further addressed in Sec.V.A.
\section{Nonperturbative 'running' in $^1S_0$ and uncoupled $P$ channels}
\subsection{$^1S_0$ channel at orders $\mathcal{O}(Q^0)$ and $\mathcal{O}(Q^2)$}Let us warm up with these two orders\cite{YH}. At order $\mathcal{O}(Q^0)$, one could see from either Eq.(\ref{relation1s0-0}) or the $T$ matrix itself\bea\frac{1}{T(p)}=J_0+i\frac{M_Np}{4\pi}+\frac{1}{C_0}\eea that there is only one constraint from scattering length: $a^{-1}=\frac {4\pi}{M_N}\left(C^{-1}_0+J_0\right)$. That is, only at this order and in $^1S_0$ channel, $J_0(\mu)$ is a running parameter, then $C_0(\mu)$ runs as below\bea\label{KSW1,KSW2} C_0(\mu)=\frac{1}{\frac{M_N}{4\pi}a^{-1}-J_0(\mu)},\eea which is exactly the KSW running for $C_0$. This is because there is only one parameter $J_0$ from loop integral and hence could be readily matched by the coupling $C_0$ in $^1S_0$ channel within the lowest order. However, it is no longer true in higher channels and/or at higher truncation orders, provided the closed-form $T$ matrices are concerned. Thus, the nonperturbative running couplings begin to show up from truncation order two.

To proceed at order $\mathcal{O}(Q^2)$, we introduce the renormalization group invariant ratios\bea\alpha_0\equiv\frac{D_{S;0}}{N_{S;0}},\ \alpha_1\equiv\frac{D_{S;1}} {N_{S;0}}.\eea Then Eqs.(\ref{relation1s0-2}) and the $T$ matrix become the following\bea&&1+\alpha_1J_3=N^{-1}_{S;0},\\\label{relations2T}&&\alpha_0J_3+\alpha_1J_5=[C_0J_3+2C_2J_5] N^{-1}_{S;0},\\& &\frac{1}{T(p)}=J_0+i\frac{M_Np}{4\pi}+\frac{1}{\alpha_0+\alpha_1p^2},\eea where it is obvious that $J_0$ is renormalization group invariant while $J_3$ and $J_5$ are running parameters and thus denoted as $J_3(\mu)$ and $J_5(\mu)$ henceforth within this order\cite{YH}. Combined with the expression of $N_{S;0}$ listed in Sec. II, we have\bea N_{S;0}=\frac{1} {1+\alpha_1J_3(\mu)}=\left[1-C_2J_3(\mu)\right]^2,\eea from which the running $C_2(\mu)$ is easy to obtain:\bea C_{2\pm}(\mu)=\left(1\pm\theta_{S}^{-\frac{1}{2}}(\mu)\right)\frac{1} {J_3(\mu)},\ \theta_{S}(\mu)\equiv N_{S;0}^{-1}=1+\alpha_1J_3(\mu).\eea From the 'boundary condition' for $C_2$: $\left.C_2\right|_{J_{3,5}=0}={\scriptstyle\frac{1}{2}}\alpha_1,$ we have\bea C_{2}(\mu)=\left(1-\theta_{S}^{-\frac{1}{2}}(\mu)\right)\frac{1}{J_3(\mu)}=\left[1-\left(1-\frac{\alpha_1}{\theta_{S}(\mu)}J_3(\mu)\right)^{\frac{1}{2}}\right] \frac{1}{J_3(\mu)}.\eea

Finally, we could find the following running coupling $C_0(\mu)$ from Eq.(\ref{relations2T}):\bea C_{0}(\mu)=\frac{\alpha_0}{\theta_{S}(\mu)}-\left[1-\left(1-\frac{\alpha_1}{\theta_{S} (\mu)}J_3(\mu)\right)^{\frac{1}{2}} \right]^2 \frac{J_5(\mu)}{J_3^2(\mu)}.\eea

Parameterizing $[J_{2n+1},n\geq1]$ in terms of a single running scale '$\mu$' as $$J_{2n+1}=\tilde{j}_{2n+1}\frac{M_N}{4\pi}\mu^{2n+1}$$ with $\tilde{j}_{2n+1}$ dimensionless, it is easy to see that the infrared and ultra-violet fixed points of such nonperturbative running couplings are\cite{YH}:\bea&&C^{(IR)}_0=\alpha_0,\ C^{(IR)}_2={\scriptstyle\frac{1}{2}} \alpha_1,\\&&C^{(UV)}_0=0,\ C^{(UV)}_2=0.\eea

If "$\tilde{j}_3\alpha_1$" is negative, then the factor $\theta_{S}^{-\frac{1}{2}}(\mu)$ in $C_2$ and $C_0$ would develop a singularity at a finite value of $\mu$, implying that the effective field theory description breaks down at that scale. This phenomenon is quite generic at higher truncation orders, see Sec. IV.B. It is actually consistent with the fact that pionless effective field theory breaks down beyond the scale of pion mass by definition. This could also happen in coupled channels\cite{AOP}. So, the ultra-violet fixed points obtained here are not trustworthy, they are even problematic, i.e., divergent at higher orders, see Sec. IV.B. below.
\subsection{$^1S_0$ channel at order $\mathcal{O}(Q^4)$} Now we consider the order $\mathcal{O}(Q^4)$ where things become more complicated. In similar fashion, we introduce the following notations for renormalization group invariant ratios\bea\beta_i\equiv\frac{N_{S;i}}{N_{S;0}},\ \alpha_i\equiv\frac{D_{S;i}}{N_{S;0}},\eea with which Eqs.(\ref{relation1s0-4}) and the $T$ matrix become\bea\label{C0relation1}&&\beta_2+\alpha_3J_3=0,\ \beta_1+\alpha_2J_3+\alpha_3J_5=0,\nonumber\\&&1+\alpha_1J_3+\alpha_2J_5+\alpha_3J_7=\frac{1}{N_{S;0}},\\ \label{C0relation2}&&\alpha_0J_3+\alpha_1J_5+\alpha_2J_7+\alpha_3J_9=\frac{C_0J_3+2C_2J_5+(2C_4+\tilde{C}_4)J_7}{N_{S;0}},\\&&\frac{1}{T(p)}=J_0+i\frac{M_Np}{4\pi} +\frac{1+\beta_1p^2+\beta_2p^4}{\alpha_0+\alpha_1p^2+\alpha_2p^4+\alpha_3p^6}.\eea Now it is clear that at this order, $J_0$, $J_3$ and $J_5$ are all renormalization group invariants with:\bea J_3=-\frac{\beta_2}{\alpha_3},\ J_5=\frac{\alpha_2\beta_2-\beta_1\alpha_3}{\alpha_3^2},\eea while $J_7,J_9$ are running parameters and will be denoted as $J_7(\mu),J_9(\mu)$ henceforth in this subsection.

To find the running couplings, let us start with $C_4(\mu)$ which is the easiest job by using the expression of $D_{S;3}$ (\ref{appA}) and $D_{S;3}=\alpha_3N_{S;0}$, the result reads\bea&&C_4(\mu)=\pm\alpha_3\left[\beta_2\theta_{S}(\mu)\right]^{-\frac{1}{2}},\nonumber\\&&\theta_{S}(\mu)\equiv N_{S;0}^{-1}=1+\alpha_1J_3+\alpha_2J_5+\alpha_3J_7(\mu).\eea Similarly, the factor $\theta^{-\frac{1}{2}}_S(\mu)$ in $C_4(\mu)$ develop a singularity provided $j_7\alpha_3<0$ (it is reasonable to suppose that $1+\alpha_1J_3+\alpha_2J_5$ is positive) so that $\theta_S=0$ around some finite value of $\mu$, which signals the breakdown of effective field theory description beyond that scale. For later convenience, we introduce the following notation for $C_4(\mu)$:\bea C_4(\mu)=s\alpha_3\left[\beta_2\theta_S(\mu)\right]^{-\frac{1}{2}},\ s^2=1.\eea

For the rest of the couplings, it is convenient to proceed in the order of $\tilde {C}_4$, then $C_2$ and finally $C_0$. After the elimination of $C_0$ and $C_2$ in the factors $[N_{\cdots}]$ and $[D_{\cdots}]$, we find that\bea&& \tilde{C}_{4\pm}=-\frac{\beta_1}{\eta}+2J_3\frac{\alpha_3\gamma\pm\left(\beta_2\zeta\right)^{\frac{1}{2}}}{\eta^2},\eea with $\gamma, \eta$ and $\zeta$ being functions of $[\alpha_{\cdots},\beta_{\cdots}]$ given in \ref{appD}. In the limits $[J_{\cdots}\Rightarrow0]$, we find that $\tilde{C}_{4\pm}\Rightarrow\alpha_2\pm2C_4,$ then only the negative sign is compatible with the following 'boundary conditions': $\alpha_0\Rightarrow C_0,\ \alpha_1\Rightarrow 2C_2,\ \alpha_2\Rightarrow2C_4+\tilde{C}_4,\ \alpha_3 \Rightarrow0.$ So, in the following solutions of $C_2$ and $C_0$, we will use\bea&&\tilde{C}_{4}=-\frac{\beta_1\eta+2\left[\beta_2\gamma+J_3\left(\beta_2\zeta\right)^{\frac{1}{2}} \right]}{\eta^2}=\frac{\alpha_3}{\beta_2}\Phi_4,\eea with $\Phi_4$ also given in \ref{appE}. To us surprise, $\tilde{C}_4$ is renormalization group invariant, an intricate point impossible to see without exploiting the tight constraints.

With $C_4(\mu)$ and $\tilde{C}_4$ given above, $C_2$ is then obtained as\bea C_2(\mu)=&&-\frac{\alpha_3}{\beta_2}+\frac{s\alpha_3}{\sqrt{\beta_2\theta_{S}(\mu)}}\left\{\frac{\alpha_2} {\alpha_3}-\frac{\beta_1}{2\beta_2}-\frac{\Phi_4}{2}\left[\frac{\alpha_3}{\beta_2}J_7(\mu)\right.\right.\nonumber\\&&\left.\left.+\left(\frac{\alpha_2}{\alpha_3}-\frac{\beta_1+s\sqrt {\beta_2\theta_S(\mu)}}{\beta_2}\right)^2\right]\right\}.\eea Finally, with $C_4(\mu),\tilde{C}_4$ and $C_2(\mu)$ given above, $C_0$ could be simply solved using Eq.(\ref{C0relation2}) \bea C_0(\mu)=&&\frac{\alpha_0}{\theta_{S}(\mu)}+\left(\frac{\beta_1}{\beta_2}-\frac{\alpha_2}{\alpha_3}\right)\left[\frac{\alpha_1}{\theta_{S}(\mu)}-\frac{2\alpha_3}{\beta_2}+\frac {s\alpha_3}{\sqrt{\beta_2\theta_S(\mu)}}\left(\frac{2\alpha_2}{\alpha_3}-\frac{\beta_1}{\beta_2}\right.\right.\nonumber\\&&\left.\left.-\left(\frac{\alpha_2}{\alpha_3}-\frac{\beta_1 +s\sqrt{\beta_2\theta_S(\mu)}}{\beta_2}\right)^2\Phi_4\right)\right]+\frac{\alpha_3}{\beta_2}\left[\frac{\alpha_3}{\beta_2}\Phi_4-\frac{\alpha_2}{\theta_S(\mu)}\right.\nonumber\\&& \left.+\frac{s\alpha_3\left(2-\left(\frac{\beta_1}{\beta_2}-\frac{\alpha_2}{\alpha_3}\right)\Phi_4\right)}{\sqrt{\beta_2\theta_S(\mu)}}\right]J_7(\mu)-\frac{\alpha^2_3}{\beta_2\theta_S (\mu)}J_9(\mu).\eea Again, the presence of the same factor $\theta^{-\frac{1}{2}}_S(\mu)$ in $C_2(\mu),C_0(\mu)$ means that these running couplings suffer from the same probable singularity as $C_4(\mu)$ does. Thus it is not an 'accident' for $C_4(\mu)$ but true for all the contact couplings, which is actually compatible with the anticipation that pionless effective field theory fails by definition beyond the scale of pion mass.

As a consistent check, we also verified that the running couplings truncated at the order $\mathcal{O}(Q^4)$ reproduce the $\mathcal{O}(Q^2)$ ones by taking $\alpha_3\rightarrow0$ and $\alpha_2\rightarrow0$, see \ref{appD}.

From the above running couplings, it is straightforward to read off the infrared and ultra-violet fixed points qualitatively:\bea&&C^{(IR)}_0=finite,\ C^{(IR)}_2=finite,\ C^{(IR)}_4=finite,\\&& C^{(UV)}_0=\infty,\ C^{(UV)}_2=\infty,\ C^{(UV)} _4=0.\eea It is clear that the ultra-violet fixed points at this order are problematic and not trustworthy as noted above, in other words, it doe not make sense to let $\mu$ go to infinity. Obviously, the remarks made at order $\mathcal{O}(Q^2)$ also apply here.

Some remarks are in order:

A). It does not make sense to ultra-violet extrapolate when there is at least one probable singularity (the position will be denoted as $\mu_{sing}$) in the running couplings beyond the order $\mathcal{O}(Q^0)$ via the 'universal' factor $\sqrt{1/\theta_S(\mu)}$. The running couplings turn into complex ones as $\mu$ goes beyond $\mu_{sing}$, where the effective field theory should be inconsistent. Also the problematic ultra-violet fixed points of $C_0$ and $C_2$ warn us against ultra-violet extrapolation. That means, the effective field theory only makes sense in a very narrow window $\mu\in[0,\mu_{sing})$. Plugging in the power counting to be specified in Sec. V.A, it is easy to see from $\theta_S(\mu)=0$ at order $\mathcal{O}(Q^2)$ that the possible pole is located at $\alpha_1J_3=-1$, thus $$\mu_{sing}=\left(-\tilde{\alpha}_1\tilde{j}_3\right)^{-1/3}\Lambda_{\not\pi}\sim2^{-1/3} \Lambda_{\not\pi},\ \alpha_1=2\frac{4\pi\tilde{\alpha}_1}{M_N}\Lambda_{\not\pi}^{-3},\ \tilde{\alpha}_1\sim\tilde{j}_3\sim\mathcal{O}(1).$$ Similar result could also be derived at order $\mathcal{O}(Q^2)$. This is in perfect accordance with the fact that effective field theory is applicable in a limited range. As far as we know, such informative contents have not been appreciated in literature yet. This is quite understandable as they obviously stem from the tight constraints that carry these information in encrypted forms, which are simply missed as they are circumvented one way or the other.

B). Note that the running behaviors presented above follow from the assumption that $J_{2n+1}\sim\frac{M}{4\pi}\mu^{2n+1}.$ In other assumptions the running behaviors would evidently differ. Thus, the running behaviors in nonperturbative contexts depend nontrivially on specific prescription in use, in contrast to perturbative cases. This is in contrast to the ones based on PDS prescription\cite{KSW1,KSW2}, which overkills the nontrivial parameters $[J_{2n+1}, n>0]$ that turn out to be the true sources of running in the closed-form $T$ matrices.

C). It is clear that the tight constraints at higher orders stop the running of 'lower' parameters in $[J_{\cdots}]$ and turn them in physical or renormalization group invariant parameters to be determined separately, while the 'highest' ones of $[J_{\cdots}]$ run. This is definitely a novel feature of pionless effective field theory (perhaps true for all effective field theory with similar truncations) in nonperturbative regimes. For example, at order $\mathcal{O}(Q^2)$, the physical or renormalization group invariant parameter is the 'lowest' one, $J_0$, while $J_3$ and $J_5$ would become renormalization group invariant beyond this order. Again, such scenario has not been considered in most literature that circumvent the tight constraints one way or the other.

D). A more pleasing and distinctive feature of our approach is that the running couplings of lower order of truncation (corresponding to lower energies or longer distances) are dependent upon the parameters ($[\alpha_{\cdots};J_{\cdots}]$) governing interaction at higher order of truncation (corresponding to higher energies or shorter-distances), not the other way around. This is in perfect accordance with the general anticipation that, the (UV) renormalization of shorter-distance interactions should be independent of the long-distance processes, while the (UV) renormalization of longer-distance interactions could be affected by the short-distance details. This is true in any channel at any given order of truncation as could be readily read off from the explicit expressions listed in Sec. IV.A and Sec. IV.B. As a further rationale, we note that such 'directional' dependence naturally accommodates the 'decoupling procedure', by which we mean the procedure of scaling down so that the higher order parameters tend to vanish ('decouple') to smoothly go over to the lower order results. This could also serve as a consistency check of the correctness of our calculations, an instance is given in \ref{appD}. This is in sheer contrast to that in KSW scheme and the like where the situation is reversed, higher order couplings are affected by the lower order ones.
\subsection{Contrast with the KSW scheme}Here we compare our nonperturbative 'running' with that obtained in Ref.\cite{KSW1,KSW2}. The latter is obtained through Taylor expansion of $Re\{1/T\}$ in term of $p^2$ in PDS. Take $^1S_0$ channel for example, $1/C_0$ conspires with $J_0$ to produce the scattering length $a$. As $J_0$ is unconstrained in such scheme, we are led to the well-known KSW running of $C_0$ depicted in Eq.(\ref{KSW1,KSW2}). The rest of the contact couplings  $[C_2,\cdots]$ also 'run' according to the understanding of the expansion in Ref.\cite{KSW1,KSW2}.

Here, we could only recover the KSW running for $C_0$ at the lowest truncation order. Once going over to higher truncations orders, the KSW running is lost as long as the closed-form is kept intact, no matter what prescription one uses. For example, working with closed-form $T$-matrices in PDS, the only superficially prescription-dependent parameter $J_0$ 'stands' alone in the inverted $T$-matrices and could not mix with the couplings if the closed-form is kept intact, i.e., not further manipulated:\bea\frac{1}{T_{^1S_0}(p)}=J_0+i\frac{M_N}{4\pi}p+\frac{1}{C_0+2C_2p^2+\cdots}.\eea That means it must be determined otherwise in principle and there is no running at all in such circumstances. Of course, different values of $J_0$ and the couplings would describe different 'physics'\cite{EPL,AOP}, for an illustration, see Sec. V.B. below.

In short, our approach has led us to an alternative and yet natural solution or scenario to effective field theories with truncations in addition to the KSW scheme and the like, which could at least serve as a supplementary choice.

Moreover, PDS is a very special prescription in that it is least informative as most of the information derived above in a general parametrization is lost: The linear relations degenerate into one trivial line: $N_{L}=1$ for any uncoupled $L$ channel. And the four remarks presented above would also be gone: A'). No trace of the possible $\mu_{sing}$; B'). No running $[J_{\cdots}]$; C'). Only one renormalization group invariant, $J_0$; D'). No sign of 'directional' relations between the interactions at different orders. The drastic change of the informational contents implies that prescription does matter to certain degree due to the tight constraints, a more general parametrization is more informative.
\subsection{Uncoupled $P$ channels at order $\mathcal{O}(Q^2)$}Here, we present the results for the uncoupled $P$ channels at order $\mathcal{O}(Q^2)$. Introducing the notation\bea \alpha_{P;0}\equiv\frac{D_{P;0}}{N_{P;0}},\eea the Eqs.(\ref{relationP2}) and $T$ matrix become\bea&&1+\alpha_{P;0}J_3=\theta_{P}(\mu),\ \alpha_{P:0}=C_{P;2}\theta_{P}(\mu), \nonumber\\& &\frac{1}{T(p)}=J_0+i\frac{M_Np}{4\pi}+\frac{1}{\alpha_{P;0}p^2}.\eea Now it is clear that $J_0$ is renormalization group invariant, but $J_3$ is not constrained and hence 'runs', then it is obvious that\bea C_{P:2}(\mu)=\frac{\alpha_{P;0}}{1+\alpha_{P;0}J_3(\mu)}=\frac{1}{\alpha_{P;0}^{-1}+J_3(\mu)}.\eea The functional form looks similar to the case of $^1S_0$ channel truncated at order $\mathcal{O}(Q^0)$: $C_0\Leftrightarrow C_{P;2}$. It is easy to read off the infrared and ultra-violet fixed points\bea C^{(IR)}_{P;2}=\alpha_{P;0},\ C^{(UV)}_{P;2}=0.\eea At this order, there might be a pole in $C_{P;2}$ at finite $\mu$ provided $\alpha_{P;0}J_3<0$.
\subsection{Uncoupled $P$ channels at order $\mathcal{O}(Q^4)$}Again, we first introduce the notations\bea\alpha_{P;i}\equiv\frac{D_{P;i}}{N_{P;0}},\ \beta_{P;i}\equiv\frac{N_{P;i}} {N_{P;0}},\eea then we have from Eqs.(\ref{NDrelationP})\bea&&\beta_{P;1}+\alpha_{P;1}J_3=0,\nonumber\\&&1+\alpha_{P;0}J_3+\alpha_{P;1}J_5=\theta_P(\mu),\nonumber\\&&\alpha_{P;0}J_5 +\alpha_{P;1}J_7=[C_{P;2}J_5+2C_{P;4}J_7]\theta_P(\mu),\\&&\frac{1}{T(p)}=J_0+i\frac{M_Np}{4\pi}+\frac{1+\beta_{P;1}p^2}{\alpha_{P;0} p^2+\alpha_{P;1}p^4}.\eea From these equations we see that, $J_0$ and $J_3$ are renormalization group invariant, while $J_5$ and $J_7$ 'run'. Then we could find the following running couplings in nonperturbative regime after imposing similar boundary conditions\bea C_{P:4}(\mu)&&=\left[1-\left(1-\frac{\alpha_{P;1}}{\theta_{P}(\mu)}J_5(\mu)\right)^{\frac{1}{2}}\right]\frac{1}{J_5(\mu)},\\C_{P:2}(\mu)&&= \frac{\alpha_{P;0}}{\theta_{P}(\mu)}-\left[1-\left(1-\frac{\alpha_{P;1}}{\theta_{P}(\mu)}J_5(\mu)\right)^{\frac{1}{2}}\right]^2 \frac{J_7(\mu)}{J^2_5(\mu)}.\eea The infrared and ultra-violet fixed points can be found from the foregoing expressions\bea&&C^{(IR)}_{P;2}=\alpha_{P;0},\ C^{(IR)}_{P;4}={\scriptstyle\frac{1}{2}}\alpha_{P;1},\\& &C^{(UV)}_{P;2}=0,\ C^{(UV)}_{P;4}=0.\eea Note again the interesting similarity in the functional forms of the nonperturbative running couplings between the $^1S_0$ channel and $P$ channels at proportionate truncation orders: $C_{0}\Leftrightarrow C_{P;2},\ C_{2}\Leftrightarrow C_{P:4}.$ We speculate that this similarity may persist at higher truncation orders. So the remarks made above in Sec. IV.B. also apply for the uncoupled $P$ channels.
\section{Power counting, Phase shifts, Wilsonian RGE and tight constraints}
\subsection{Power counting and truncation}At this point, one may ask how power counting are manifested in our approach. The answer lies in the renormalization group invariant ratios $[N_{\cdots}/N_{\cdots;0},D_{\cdots}/N_{\cdots;0}]$ or $[\beta_{\cdots},\alpha_{\cdots}]$ that parameterize the closed-form $T$ matrices. That means, for a power counting to be meaningful for an effective field theory, it must manifest itself in the renormalized objects. As a very naive guess, we may have\bea\label{EFTPC}|\beta_{L;i}|\sim \Lambda_{\not\pi}^{-2i},\ |\alpha_{L;j}|\sim\eta\frac{4\pi}{M_N}\Lambda^{-2j-2L-1}_{\not\pi},\nonumber\\ \forall i>0,\ j\geq0,\ \eta=1(\texttt{\small diagonal}),2(\texttt{\small off-diagonal}). \eea With such a power counting and noting that the renormalization group invariant $J_0$ could be counted as $J_0\sim|\alpha^{-1}_{S;0}|\sim\frac{M_N}{4\pi}\Lambda_{\not\pi}$, one could arrive at a large scattering length provided $J_0$ and $\alpha_{S;0}$ are of the opposite sign, namely,\bea a^{-1}=-\frac{4\pi}{M_N}\left(J_0+\alpha^{-1}_{S;0}\right)\sim -\mathcal{O}(\Lambda_{\not\pi}) +\mathcal{O}(\Lambda_{\not\pi})\sim\mathcal{O}(\epsilon\Lambda_{\not\pi}),\ \epsilon\ll1.\eea This is natural to achieve in contrast to the KSW scheme that suffers from extra large ERE form factors and other problems, see Refs.\cite{Cohen,FMS} and references therein. In other words, it is the cooperation of the renormalization group invariant ratios AND the renormalization group invariant parameter $J_0$ determine whether the scattering length is unnaturally large or not. Thus we may simply work a natural power counting for the ratios $[\beta_{\cdots},\alpha_{\cdots}]$ that may well accommodate both natural and unnatural scenarios for $NN$ scattering by appropriate choice of $J_0$ (more complicated situations may be achieved as more renormalization group invariant parameters $[J_{\cdots}^{\texttt{\tiny(phys)}}]$ are available).

It is clear from our presentation that at each order of truncation, the parameters $[J_{2k+1},k>0]$ with lower mass dimensions become 'unmatched' with effective field theory couplings and hence renormalization group invariants, the ones with highest mass dimensions 'run'. Thus in the limit that $\Delta\rightarrow\infty$, all $[J_{\cdots}]$ would become renormalization group invariants, and the running couplings tend to be renormalization group invariants, too. This is not a surprise, however, from the underlying theory perspective. As $\Delta\rightarrow\infty$, the truncation tends to be removed completely, so a complete description would be recovered. This is a general claim from the effective field theory/underlying theory duality perspective that should be applicable to all consistent field theories.

It is natural to see that the effective field theory power counting rules delineated in Eq.(\ref{EFTPC}) automatically allow for corrections/adjustments of the renormalization group invariant ratios $[\beta_{\cdots},\alpha_{\cdots}]$ in a systematic manner, in order to yield physical ERE factors. At a given order of effective field theory truncation $\Delta$, the differences between the exact values and the theoretical values of these ratios are of higher order\bea\frac{\beta_{L;i}^{(\Delta)}-\beta_{L;i}^{(\texttt{\tiny phys})}} {\beta_{L;i}^{(\Delta)}}=\mathcal{O}\left(\frac{Q^{\delta}}{\Lambda_{\not\pi}^{\delta}}\right),\quad\frac{\alpha_{L;j}^{(\Delta)}-\alpha_{L;j}^{(\texttt{\tiny phys})}} {\alpha_{L;i}^{(\Delta)}}=\mathcal{O}\left(\frac{Q^{\delta^\prime}}{\Lambda_{\not\pi}^{\delta^\prime}}\right),\quad\delta\geq1,\ \delta^\prime\geq1.\eea Of course, these power counting rules could be readily transcribed into the effective field theory couplings in terms of prescription independence of the ratios or their algebraic combinations, i.e., the physical ERE parameters. We should note that these renormalization group equations are derived in exactly the standard fashion where prescription independence of certain objects is the starting point, involving no additional assumptions.

So far, our discussions are limited to the simpler pionless effective field theory. The realistic situations with pion exchanges would make the running behaviors more complicated. For example, rather than simple fixed points, there might be limit cycles\cite{limitcycleWil,limitcycle1,limitcycle2}. In our presentation, the complicated running behaviors might seem to stem from truncations in nonperturbative contexts. In this connection, we note that the extra divergences or parameters that arise in effective field theory as truncations do not commute with loop integrations, must correspond to some well-defined quantities if one could calculate with underlying theory. Therefore, the tight constraints and the running behaviors thus derived must have reflected at least part of the 'truths' provided the effective field theory is a rational one.
\subsection{Phase shifts of $^1S_0$ from closed-form $T$ matrix}In Fig. 1, we plotted the phase shift curves 'predicted' by the on-shell closed-form $T$ matrices in the $^1S_0$ channel with the PDS prescription for simplicity (where the expressions are greatly simplified and $\tilde{C}_4$ can be merged into $C_4$ in the $T$ matrix) in Fig.1. The couplings are determined from ERE parameters under two choices of $J_0$, see Table 1. Obviously, the phase shifts 'predicted' improve with the order of truncation, and the '$\frac{4\pi}{M_N}J_0=138$' choice (right) out behave the '$\frac{4\pi}{M_N}J_0=35$' choice (left), showing that $J_0$ is 'physical' in the closed-form $T$ matrix or tight constraints do matter. In most literature, tight constraints are 'removed' by expanding the closed-form $1/T$ beyond the leading order, 'reproducing' the effective range theory (ERT)\cite{KSW1,KSW2}. But such 'equivalence' to ERT is achieved by 'further manipulation' on the closed-form $T$, hence a 'fake equivalence'.
\begin{figure}[t]\begin{center}\label{Fig1}
\resizebox{16cm}{!}{\includegraphics{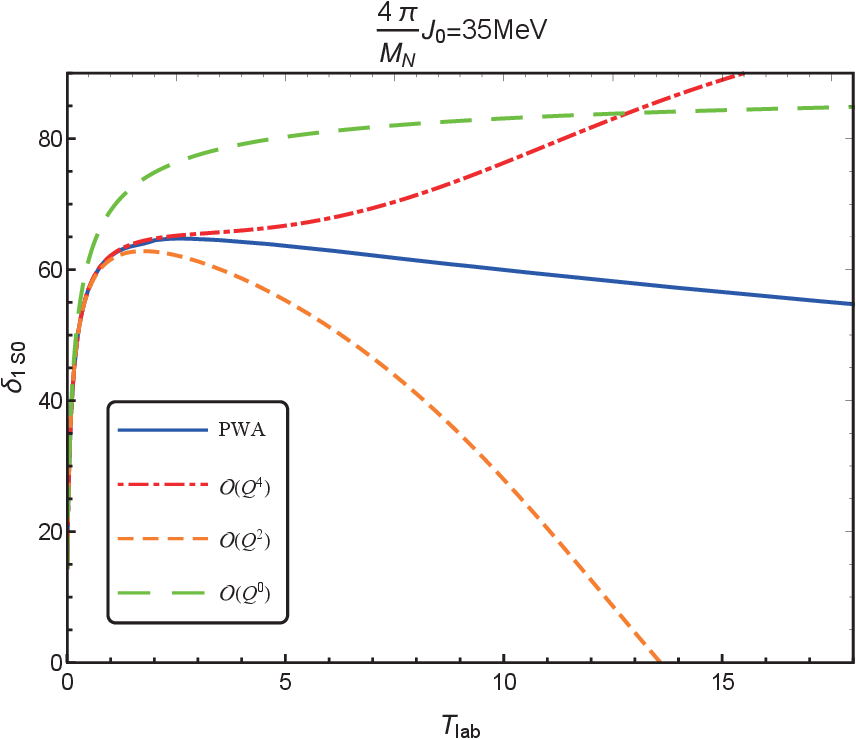}\quad\quad\quad\includegraphics{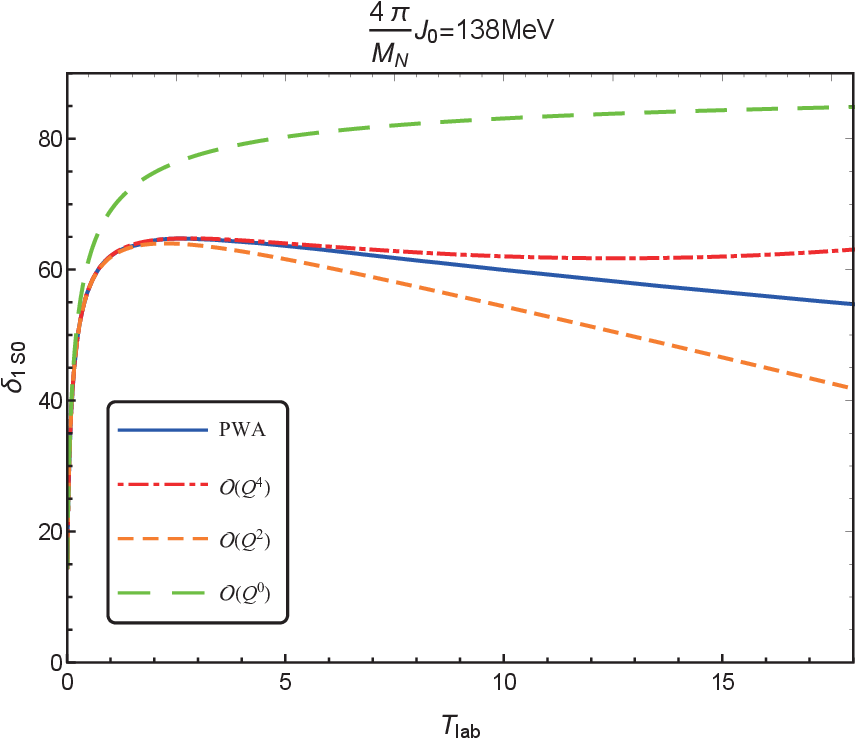}}\caption
{Phase shifts for $^1S_0$ NN scattering: Nijmegen data (PWA) versus predictions from closed-form $T$ matrices at orders $Q^0,Q^2,Q^4$. The abscissa $T_{\text{\tiny lab}}$ denotes the energy in laboratory frame, unit in MeV.}\end{center}\end{figure}
\begin{table}[ht]\caption{Couplings from ERT}
\begin{center}\begin{tabular}{c|ccc}\hline\hline\\$\frac{4\pi}{M_N}J_0$[MeV]\quad\quad&\quad$C_0$[GeV$^{-2}$]\quad&\quad$C_2$[GeV$^{-4}$]\quad&\quad$C_4$[GeV$^{-6}$]\quad\\\\\hline \\35&$-3.0898\times10^2$&$2.4222\times10^4$&$-4.0205\times10^6$\\\\138&$-9.1467\times10^1$&$2.1226\times10^3$&$-1.1804\times10^5$\\\\\hline\hline
\end{tabular}\end{center}\end{table}
\begin{table}[ht]\label{ERTscale}\caption{Scales involved in Table 1}
\begin{center}\begin{tabular}{c|ccc}\hline\hline\\$\frac{4\pi}{M_N}J_0[$MeV$]$&\quad$\Lambda_0[$MeV$]$\quad&\quad$\Lambda_2[$MeV$]$\quad&\quad$\Lambda_4[$MeV$]$\quad\\\\ \hline\\35&43.3$\sim$(0.31$m_\pi$)&294.5$\sim$(2.13$m_\pi$)&286.2$\sim$(2.07$m_\pi$)\\\\138&146.3$\sim$(1.06$m_\pi$)&184.7$\sim$(1.34$m_\pi$)&186.7$\sim$(1.35$m_\pi$)\\\\\hline\hline
\end{tabular}\end{center}\end{table}
To examine the power counting, we extract the scales involved as below:\bea\frac{4\pi}{M_N}J_0=35:&&C_0=\frac{4\pi}{M_N}\Lambda_0^{-1},\ C_2=\frac{4\pi}{M_N}\Lambda^{-1}_2\Lambda^{-2}_0,\ C_4=\frac{4\pi}{M_N}\Lambda_4^{-2}\Lambda^{-3}_0,\\\frac{4\pi}{M_N}J_0=138:&& C_0=\frac{4\pi}{M_N}\Lambda_0^{-1},\ C_2=\frac{4\pi}{M_N}\Lambda^{-3}_2,\ C_4=\frac{4\pi}{M_N}\Lambda_4^{-5}.\eea The results are listed in Table 2. The first row serves to mimic the KSW scaling as $\frac{4\pi}{M_N}J_0=35$MeV is the typical value of a running scale, where the 'large' scales extracted are much larger than the upper scale: $\Lambda_2\sim\Lambda_4>2\Lambda_{\not\pi}>6\Lambda_0$, where we take $\Lambda_{\not\pi}=m_\pi$. While with $\frac{4\pi}{M_N}J_0=138$MeV, a natural scaling (c.f. Sec.V.A) seems to work better: $\Lambda_0\sim\Lambda_2\sim\Lambda_4\sim\Lambda_{\not\pi}$.
\subsection{Wilsonian RGE, tight constraints and power counting}Before closing this section, we wish to issue the following remarks: First, the primary distinction between our renormalization group equations and others is that the tight constraints are intensively exploited with a general parametrization of loop integrals in demanding that physical properties or observables be insensitive to variations in prescription. In Wilsonian or exact renormalization group equations\cite{Birse,Naka,Harada}, there is no room for more sophisticated situations (Sec. III.B), hence one may be led to flawed judgement about effective field theory power counting\cite{EGeM}; Second, therefore, the most natural starting point for deriving power counting is simply the basic idea of low energy expansion of a fundamental field theory in terms of local interactions in terms of low energy effective degrees, and we would generically arrive at a natural power counting. Both natural and unnatural low energy behaviors could be well achieved with the help of parameters like $J_0$ in accordance with tight constraints. Third, a general parametrization would actually allow us to accommodate the St\"ukelberg-Peterman form of renormalization group equation\cite{Stuek}. Fourth, as is shown in Sec. IV.B, the (probable) singularities in the nonperturbative running couplings mean that the effective theory is only consistent within a limited window. In contrast, as the sliding scale is not limited at all, it is somehow inconsistent to apply Wilsonian or exact renormalization group equation to the effective field theories that only make sense below a finite scale. Therefore, our presentations above at least provide an important conceptual supplementary to renormalization group equation in effective field theory contexts.
\section{Summary}In this report, we have examined the running behaviors of the contact couplings of pionless effective field theory in nonperturbative contexts for $^1S_0$ and uncoupled $P$ channels. Starting with the closed-form $T$ matrices in a general parametrization of divergent integrals, some interesting linear relations among the factors and parameters for the $T$ matrices were presented. Working in the underlying theory perspective, we arrived at novel renormalization group equations by exploiting the tight constraints imposed by the closed-form $T$ matrices and obtained the running couplings that exhibit probable singularities. Then we demonstrated that the novel renormalization group equations and their solutions could be quite informative about the physics delineated by an effective field theory and its underlying theory provided that the tight constraints imposed by the closed-form $T$-matrices are adequately exploited than discarded. Brief comparisons with other literature were also presented.

We also wish to note the following distinctive aspects about our study of renormalization group equations: 1). The starting point is the underlying theory perspective in the sense of being UV complete. 2). A general parametrization of the loop integrals in effective field theory is employed and proves to be crucial. 3). The tight constraints are exploited rather than circumvented, making certain parameters 'physical'. 4). The renormalization group equations are derived within the circumstances of tight constraints. 5). The running couplings are tractable from such tightly constrained renormalization group equations. 6). More (globally) structural and self-consistent properties of the effective field theory are encrypted into the running couplings due to the tight constraints. 7). It is shown that Wilsonian renormalization group equation and the like is incompatible with tight constraints due to rigidity of cutoff regulator, hence probably problematic for issues with tight constraints. 8). The power counting should be naturally derived right from the basic idea of effective field theory rather than from renormalization group equations.

For most nonperturbative problems, a relativistic or covariant framework are very difficult to come by, one is often forced to work with certain non-relativistic expansion and other truncations and then encounters tight constraints of various forms. Our discussions here amounts to pointing out an alternative way to make sense of these constraints rather than resorting to means to circumvent them at all.
\section*{Acknowledgement}Our gratitude to an anonymous referee for his/her comments is happily acknowledged. This project is supported in part by the National Natural Science Foundation of China under Grant No. 11435005 and by the Ministry of Education of China.
\appendix
\section{$N$'s and $D$'s for $^1S_0$ channel at order $\mathcal{O}(Q^4)$}\label{appA}To solve the Lippmann-Schwinger equations for $NN$ scattering with contact potentials, the following integrals would be needed\bea \label{integrals}&&\int\frac{d^3k}{(2\pi)^3}\frac{1}{E^+-\frac{k^2}{M_N}}\equiv-\mathcal{I}_0=J_0+\frac{M_N}{4\pi}ip,\\&&\int\frac{d^3k}{(2\pi)^3} \frac{k^{2n}}{E^+-\frac{k^2}{M_N}}\equiv \sum_{k=1}^nJ_{2k+1}p^{2(n-k)}-\mathcal{I}_0p^{2n},\ n\geq1.\eea Here $[J_0,J_{2k+1}]$ with $k=1,2,3,4$ are primarily prescription-dependent parameters before taking the tight constraints imposed by the closed-form $T$ matrices into account.

With the parameters defined above, the detailed expressions for $N$'s and $D$'s for $^1S_0$ channel at order $\mathcal{O}(Q^4)$ read,\bea N_{S;0}=&&(1-C_2J_3-C_4J_5)^2-C_0\tilde{C}_4 J_3^2-\tilde{C}_4J_5+2C_4\tilde{C}_4(J_5^2-J_3J_7)\nonumber\\&&-C_4^2\tilde{C}_4(J_5^3+J_3^2J_9-2J_3J_5J_7),\\N_{S;1}=&&-2C_4J_3-\tilde{C}_4J_3+2C_2C_4J_3^2+2\tilde{C}_4C_4J_3J_5 +2C_4^2J_3J_5+C_4^2\tilde{C}_4(J_3^2J_7-J_3J_5^2),\\N_{S;2}=&&C_4^2J_3^2,\\D_{S;0}=&&C_0+C_2^2J_5-C_0\tilde{C}_4J_5+2C_2C_4J_7+C_4^2J_9+C_4^2\tilde{C}_4(J_7^2-J_5J_9),\\D_{S;1} =&&2C_2-C_2^2J_3+C_0\tilde{C}_4J_3+C_4^2J_7+2C_4\tilde{C}_4J_7+C_4^2\tilde{C}_4(J_3J_9-J_5J_7),\\D_{S;2}=&&2C_4+\tilde{C}_4-2C_2C_4J_3-C_4^2J_5 -2C_4\tilde{C}_4J_5+C_4^2\tilde{C}_4(J_5^2-J_3J_7),\\D_{S;3}=&&-C_4^2J_3.\eea
\section{$N$'s and $D$'s for $P$ channels at order $\mathcal{O}(Q^4)$}\label{appB}Similarly, the factors for $P$ channels $T$ matrices at order $\mathcal{O}(Q^4)$ read,\bea&&N_{P;0} =(1-C_{P;4}J_5)^2-C_{P;2}J_3-C_{P;4}^2J_3J_7,\\&&N_{P;1}=C_{P;4}^2J_3J_5-2C_{P;4}J_3,\\&&D_{P;0}=C_{P;2}+C_{P;4}^2J_7,\\&&D_{P;1}=2C_{P;4}-C_{P;4}^2J_5.\eea
\section{Off-shell version of $T$ matrix at order $\mathcal{O}(Q^4)$}\label{appC}At this order of truncation, the off-shell $T$ matrix of $^1S_0$ channel take more involved form as below\bea T(q,q^\prime;p)&=&\frac{\tilde{D}_S(q,q^\prime;p)+\check{\delta}_S(q,q^\prime;p)\check{O}_{\texttt{\tiny off}}}{\sum_{i=0}^2N_{S;i}p^{2i}+\mathcal{I}_0\left(\sum_{j=0}^3 D_{S;j}p^{2j}\right)},\\\tilde{D}_S(q,q^\prime;p)&=&V_{^1S_0}(q,q^\prime)+\left(C^2_2-C_0\tilde{C}_4\right)J_5+2C_2C_4J_7+C^2_4J_9+\tilde{C}_4C^2_4\left(J_7^2-J_5J_9\right) \nonumber\\&&+\left[\left(C^2_2-C_0\tilde{C}_4\right)J_3+C_2C_4J_5-\tilde{C}_4C_4J_7+\tilde{C}_4C^2_4\left(J_5J_7-J_3J_9\right)\right]\nonumber\\&&\times\left(p^2-q^2-{q^\prime}^2\right) +\left[C_2C_4J_5+C_4\left(C_4+\tilde{C}_4\right)J_7\right]p^2\nonumber\\&&+\left[C_2C_4J_3+C^2_4J_5+\tilde{C}_4C^2_4\left(J_3J_7-J5^2\right)\right]\left(p^4-q^4-{q^\prime}^4\right) \nonumber\\&&+\left[C_2C_4J_3p^2+\tilde{C}_4C_4J_5\left(q^2+{q^\prime}^2\right)\right]\left(p^2-q^2-{q^\prime}^2\right)\nonumber\\&&+C_4^2J_3\left(1-\tilde{C}_4J_5\right)p^2\left( p^4-q^4-{q^\prime}^4\right)+\tilde{C}_4C^2_4J_3J_5q^2{q^\prime}^2\left(p^2-q^2-{q^\prime}^2\right),\\\check{\delta}_S(q,q^\prime;p)&=&-J_0\left\{C^2_2-\tilde{C}_4\left(C_0+C^2_4J_9 \right)+C_4\left(C_2+\tilde{C}_4C_4J_7\right)\left(p^2+q^2+{q^\prime}^2\right)+C_2C_4p^2\right.\nonumber\\&&\left.+C_4\left[C_4+\tilde{C}_4\left(1-C_4J_5\right)\right]p^2\left(q^2 +{q^\prime}^2\right)+C^2_4\left[p^4+\left(1-\tilde{C}_4J_5+\tilde{C}_4J_3p^2\right)\right.\right.\nonumber\\&&\times\left.\left.q^2{q^\prime}^2\right]\right\}+\tilde{C}_4C_4J_3 \left(q^2+{q^\prime}^2\right)+\tilde{C}_4C^2_4\left(J^2_5+J_3J_5p^2+J^2_3q^2{q^\prime}^2\right).\eea From the definitions given above, one could readily verify that $\tilde{D}_S(p,p;p)=\sum_{j=0}^3D_{S;j}p^{2j}$.

For the uncoupled $P$ channels, we have similarly\bea T(q,q^\prime;p)&=&\frac{\tilde{D}_P(q,q^\prime;p)+\check{\delta}_P(q,q^\prime;p)\check{O}_{\texttt{\tiny off}}}{\sum_{i=0}^1 N_{P;i}p^{2i}+\mathcal{I}_0\left(\sum_{j=0}^1D_{P;j}p^{2j}\right)},\\\tilde{D}_P(q,q^\prime;p)&=&V_P(q,q^\prime)+C^2_{P;4}\left[J_7+J_5\left(p^2-q^2-{q^{\prime}}^2\right)\right] qq^\prime,\\\check{\delta}_P(q,q^\prime;p)&=&\left(J_3-\mathcal{I}_0p^2\right)C^2_{P;4}qq^\prime.\eea Again, it is easy to verify that $\tilde{D}_P(p,p;p)=\left(\sum_{j=0}^1D_{P;j}p^{2j}\right)p^2$.
\section{Consistency check at order $\mathcal{O}(Q^2)$}\label{appD}In order to reproduce the order $\mathcal{O}(Q^2)$ running couplings using the $\mathcal{O}(Q^4)$ ones, we need to let $[J_7,J_9,\alpha_2,\alpha_3,\beta_1,\beta_2]$ go to zero while keeping $J_3,J_5$ finite, that means,\bea\beta_2=o(\alpha_3)J_3,\ \beta_1=o\left(\alpha_3\right)J_5,\ \alpha_2=o\left( \alpha_3\right)J^{-1}_3J_5.\eea This in turn means that as $\alpha_3=\epsilon \rightarrow0$, we have,\bea&&\eta=J_3(1+\alpha_1J_3)+o\left(\epsilon^{\frac{1}{2}}\right),\\&&\theta= {1+\alpha_1J_3}+o\left(\epsilon^{\frac{1}{2}}\right),\\&&\zeta=(1+\alpha_1J_3)^2+o\left(\epsilon^{\frac{1}{2}}\right),\eea with which we find that\bea&&C_4=o\left(\epsilon^{\frac{1} {2}}\right),\ \tilde{C}_4=o\left(\epsilon^{\frac{1}{2}}\right),\\&&C_2={J_3^{-1}}\left[1-\left({1+\alpha_1J_3}\right)^{-\frac{1}{2}}\right]+o\left(\epsilon^ {\frac{1}{2}}\right),\\&& C_0=\frac{\alpha_0J_3+\alpha_1J_5}{J_3+\alpha_1J^2_3}-\frac{2J_5}{J_3^2}\left[1-\left(1+\alpha_1J_3\right)^{-\frac{1}{2}}\right]+o\left(\epsilon^{\frac{1}{2}}\right).\eea

In going over to the lowest order, one needs to let $\alpha_1,J_3$ and $J_5$ go to zero. Then we have, $C_0=\alpha_0+o(\epsilon),\ C_2=o(\epsilon).$ Here we note the striking result that in the perspective adopted here, the $C_0$ coupling is constrained by the physical parameter $\alpha_0$ when one try to recover the lower order results from higher orders.
\section{Some definitions of renormalization group invariant combinations}\label{appE}Here we list some detailed expressions of the notations used in Sec. IV.B in $\tilde{C}_4$, $C_2$ and $C_0$:\bea&&\gamma\equiv J_5-\alpha_0J_3^2=J_3\tilde{\gamma},\ \tilde{\gamma}\equiv\frac{\beta_1}{\beta_2}+\frac{\alpha_0\beta_2-\alpha_2}{\alpha_3},\\& &\eta\equiv J_3(1+\alpha_1J_3)-\beta_1J_5=J_3\tilde{\eta},\ \tilde{\eta}\equiv1+\frac{\beta_1\alpha_2-\beta_2\alpha_1}{\alpha_3}-\frac{\beta^2_1}{\beta_2},\\&&\zeta\equiv (1+\alpha_1J_3)^2+\alpha_2J_5-\beta_1(\alpha_0\eta+\alpha_1J_5)+\alpha_0\alpha_3J_3(J_5+\gamma)\nonumber\\&&=\left(1-\frac{\alpha_1\beta_2}{\alpha_3}\right)^2+\left(\alpha_1\beta_1- \alpha_2\right)\frac{\alpha_3\beta_1-\alpha_2\beta_2}{\alpha_3^2}\nonumber\\&&+\frac{\alpha_0}{\alpha_3}\left[3\beta_1\beta_2-\beta^3_1+\beta_2\frac{\alpha_2(\beta^2_1-2\beta_2)- \alpha_1\beta_1\beta_2}{\alpha_3}\right]+\alpha_0^2\frac{\beta^3_2}{\alpha_3^2};\\&&\Phi_4\equiv\frac{\beta_1}{\tilde{\eta}}+\frac{2\beta_2\tilde{\gamma}}{\tilde{\eta}^2}+\frac{2 \left(\beta_2\zeta\right)^{\frac{1}{2}}}{\tilde{\eta}^2}.\eea

\end{document}